\documentclass{aa}
\usepackage{times}
\usepackage{amssymb}
\usepackage{graphicx}
\usepackage{natbib}

\begin{document}



\title{The near-synchronous polar \object{V1432\,Aql}
 (\object{RX\,J1940.1--1025}):
 Accretion geometry and synchronization time scale}

\author{
R. Staubert\inst{1} \and
S. Friedrich\inst{2,3} \and
K. Pottschmidt\inst{3,4} \and
S. Benlloch\inst{1} \and
S.L. Schuh\inst{1} \and
P. Kroll\inst{5} \and
E. Splittgerber\inst{5} \and
R. Rothschild\inst{6}}

\institute{
Institut f\"ur Astronomie und Astrophysik,
Universit\"at T\"ubingen,
Sand 1,
D-72076 T\"ubingen,
Germany \and
Institut f\"ur Theoretische Physik und Astrophysik,
Universit\"at Kiel,
Leibnizstr. 15,
D-24098 Kiel,
Germany \and
Max-Planck-Institut f\"ur extraterrestrische Physik,
Giessenbachstr.,
D-85748 Garching.
Germany \and
Integral Science Data Center,
16 Chemin d'Ecogia
CH-1290 Sauverny,
Switzerland \and
Sternwarte Sonneberg,
Sternwartestr. 32,
D-96515 Sonneberg,
Germany \and
Center for Astrophysics and Space Science (CASS),
Univ. California, San Diego,
California,
USA
}

\offprints{R. Staubert, staubert@astro.uni-tuebingen.de}

\titlerunning{The near-synchronous polar \object{V1432\,Aql}:
 Accretion geometry and synchronization time scale}

\authorrunning{Staubert et al.}
\date{Received $<$24.03.03$>$ / Accepted $<$02.06.03$>$}

\abstract{The magnetic Cataclysmic Variable (mCV) \object{V1432\,Aql}
  (\object{RX\,J1940.1--1025}) belongs to the four-member subclass of
  near-synchronous polars with a slight non-synchronism ($<2\,\%$)
  between the spin period of the white dwarf and the binary period. In
  these systems the accretion geometry changes periodically with phase
  of the beat cycle. We present the application of a dipole accretion 
  model for near-synchronous systems developed by Geckeler \& Staubert 
  (1997a) to extended optical and X-ray data. 
  We detect a significant secular 
  change of the white dwarf spin period in \object{V1432\,Aql} of 
  ${\rm dP}_{\rm spin}/{\rm dt}=-5.4^{+3.7}_{-3.2}\cdot10^{-9}$\,s/s 
  from the optical data set alone. This corresponds to a 
  synchronization time scale $\tau_{\rm sync}=199^{+441}_{-75}$\,yr, 
  comparable to the time scale of 170\,yr for \object{V1500\,Cyg}.
  The synchronization time scale in \object{V1432\,Aql} is in excellent
  agreement with the theoretical prediction from the dominating
  magnetic torque in near-synchronous systems. We also present period 
  analyses of optical CCD photometry and RXTE X-ray data, which argue 
  against the existence of a 4000\,s period and an interpretation of
  \object{V1432\,Aql} as an intermediate polar. The dipole accretion 
  model also allows to constrain the relevant parameters of the 
  accretion geometry in this system: the optical data allow an
  estimate of 
  the dimensionless parameter $({\rm R}_{\rm t0}'/{\rm R}_{\rm
  wd})^{1/2}\sin{\beta}=3.6_{-1.1}^{+2.7}$, with a lower limit for
  the threading radius of ${\rm R}_{\rm t0}'>10\,{\rm R}_{\rm wd}$ 
  (68\% confidence).     
  }

\maketitle

\section{Introduction}

Polars are magnetic Cataclysmic Variables consisting of a Roche
lobe-filling secondary and a white dwarf primary with a magnetic
moment $\mu\geq 10^{34}$\,G\,cm$^3$, exceeding those of neutron stars
in low mass X-ray binaries (LMXB) with $\mu\approx10^{26}$\,G\,cm$^3$
and in high mass X-ray binaries (HMXB) with
$\mu\approx10^{30}$\,G\,cm$^3$ (Cropper, 1998). The field
strengths of most white dwarfs in polars are in the range of 7--70\,MG
(Beuermann \& Burwitz, 1995). However, a
few high field polars breaking the 70\,MG barrier have been found,
e.g., \object{AR\,UMa} with a reported field strength of 230\,MG
(Schmidt et al., 1996).

The strong magnetic field of the white dwarf in polars prevents the
formation of an accretion disk and funnels the accreting matter
transferred from the secondary directly onto the white dwarf
photosphere near one or both of the magnetic poles. In addition, also
because of the strong magnetic field, the white dwarf spin period and
the orbital period are synchronized in most polars (Lamb et
al., 1983, Wu \& Wickramasinghe, 1993).

The post shock region of the accretion funnel emits hard X-ray
bremsstrahlung radiation from a plasma heated to a few $10^8$\,K (Lamb
\& Masters, 1979). A soft X-ray/EUV blackbody component with
${\rm T}_{\rm eff}$ of a few $10^5$\,K results from the reprocessing
of hard X-rays and the thermalising of the kinetic energy of dense
accretion blobs in the white dwarf photosphere (Kuijpers \&
Pringle, 1982, Frank et al., 1988, Hameury et al., 1989). 
Optical and near infrared spectra show cyclotron
emission components also originating from the post shock flow
(Chanmugam \& Langer, 1991, Wickramasinghe \&
Meggitt, 1985). They provide an important diagnostic tool for
probing the physical parameters in the accretion region and
determining the accretion geometry (Schwope, 1990, 1995, and 
Schwope et al., 1995).

In this paper we will focus on a special subclass of polars with a
slight non-synchronous rotation of the white dwarf with respect to the
binary period. In these systems the orientation of the magnetic field
of the white dwarf relative to the secondary changes continously with
phase of the beat cycle. The varying accretion geometries in those
systems and the detection of secular changes in the white dwarf spin
periods (for three of the four known near-synchronous polars) offer
unique insights into the accretion geometry and torques at work
in polars. We will present a detailed model for the accretion geometry
in near-synchronous systems and an application to our extended optical
and X-ray data set on \object{V1432\,Aql}.

\section{\object{V1432\,Aql} and the 
Near-Synchronous Polars}

There exists a small subclass of four polars, consisting of
\object{V1432\,Aql}, \object{V1500\,Cyg}, \object{BY\,Cam} (all with
orbital periods near 3.4\,h), and \object{RX\,J2115.7--5840} (${\rm
  P}_{\rm orb}\approx1.85$\,h, the first system below the 2--3\,hr
period gap, Schwope et al., 1997, Ramsay et
al., 1999), with a slight ($<2\,\%$) but significant
non-synchronous rotation of the white dwarf, see Table~\ref{nsync}.
Although the white dwarf is presumably temporarily out of
synchronization, these objects clearly do belong to the group of
`classical' polars and not to the intermediate polars, where the spin
period of the white dwarf is substantially shorter than the orbital
period of the binary system (Patterson, 1994).

\begin{table}
\caption[]{\protect\label{nsync}
Comparison of near-synchronous polars}
\begin{flushleft}
\begin{tabular}{crrc} \hline
\multicolumn{1}{c}{Object} &
\multicolumn{1}{c}{$\Delta {\rm P}/{\rm P}_{\rm orb}^{\rm a}$} &
\multicolumn{1}{c}{$\tau_{\rm sync}$ [yr]} &
\multicolumn{1}{c}{Ref.$^{\rm b}$} \\ \hline
\object{V1500\,Cyg} & --1.8\% & $\sim150$ &
(1)\\
$''$ & $''$ & 290$\pm$150 &
(2)\\
$''$ & $''$ & 170$\pm$8 &
(3)\\ \hline
\object{BY\,Cam} & --1.0\% & 1600$\pm$500 &
(4), (5)\\ \hline
\object{RX\,J2115--58}$^{\rm c}$ & $\sim-1$\% & -- &
(6)\\
$''$ & --1.2\% & -- &
(7)\\ \hline
\object{V1432\,Aql} & $\sim0.3$\% & $\sim100$ &
(8)\\
$''$ & 0.28\% & $199^{+441}_{-75}$ &
see Table~\ref{tau}\\ \hline
\end{tabular}
\end{flushleft}
a: $\Delta {\rm P}/{\rm P}_{\rm orb}=
({\rm P}_{\rm spin}-{\rm P}_{\rm orb})/{\rm P}_{\rm orb}$\\
b: References:
(1) Schmidt \& Stockman~(1991);
(2) Pavlenko \& Pelt~(1991);
(3) Schmidt et al.~(1995);
(4) Silber et al.~(1997);
(5) Mason \& Chanmugam~(1992);
(6) Schwope et al.~(1997);
(7) Ramsay et al.~(1999);
(8) Geckeler \& Staubert~(1997a).\\
c: First system below the period gap
\end{table}

Through observations with Rosat, \object{RX\,J1940.1--1025} has been
identified as the source of the periodically modu\-lated X-ray flux
which had earlier been associated with the Active Galaxy
\object{NGC\,6814}, positioned 37$'$ from the polar (Madejski et
al., 1993, Staubert et al., 1994). Rosen et
al.~(1993) identified the optical counterpart of the new X-ray
source \object{RX\,J1940.1--1025}, later named \object{V1432\,Aql}.
Strong emission lines in the optical spectrum combined with the high
X-ray flux suggested the classification of the object as a magnetic
CV. Patterson et al.~(1993) detected a variability in the
optical flux of the source with a period of $\sim12150$\,s.

Spectroscopic studies by Staubert et al.~(1993) revealed a
periodic variation of the radial velocity of the H$\alpha$ emission
line with a period of $12\,120\pm3$\,s. The authors suggested that
\object{V1432\,Aql} may be a non-synchronous polar, which has
subsequently been verified by extended optical and X-ray observations
(Staubert et al., 1994, Friedrich et al., 1996a,
Patterson et al., 1995,  Watson et al., 1995, Geckeler \&
Staubert, 1997a). Of the four known near-synchronous polars,
\object{V1432\,Aql} has the smallest separation (0.28\%) between the
orbital period and the spin period and, contrary to the three others,
the orbital period in \object{V1432\,Aql} is the shorter one, which
poses an interesting theoretical problem that is not yet understood
(J. Frank, priv. comm.).

In the near-synchronous polars the orientation of the magnetic field
of the white dwarf primary relative to the secondary changes
continously with phase $\Phi_{\rm beat}$ of the beat cycle, with the
beat period defined as ${\rm P}_{\rm beat}^{-1}= \mid {\rm P}_{\rm
  spin}^{-1}-{\rm P}_{\rm orb}^{- 1}\mid$.  The accreting matter
delivered by the secondary thus follows different field lines due to
the changing magnetic geometry and impacts on the white dwarf surface
at different positions relative to the magnetic pole, depending on the
phase of the beat cycle. Geckeler \& Staubert~(1997a) have
provided first results on the detection of this effect in the
near-synchronous system \object{V1432\,Aql}.

\begin{figure}
\resizebox{\hsize}{!}{\includegraphics{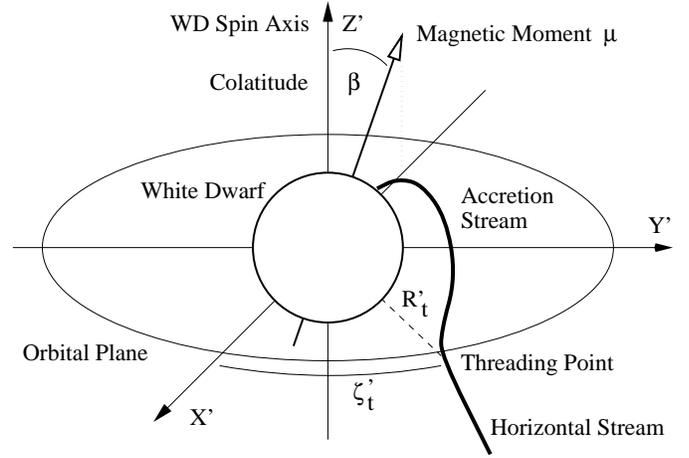}}
\caption{Diagram of the accretion geometry
in the spin system (X$'$Y$'$Z$'$) of the white dwarf,
the Z$'$-axis is the white dwarf spin axis
(Geckeler \& Staubert, 1997a).}
\label{pol}
\end{figure}

There are hints for at least two of the near-synchronous systems
that the synchronization of the white dwarf has been broken as the
result of a nova outburst. Asynchronism was observed in
\object{V1500\,Cyg} after it had erupted as \object{Nova Cygni
  1975} (Stockman et al., 1988) and spectra of \object{BY\,Cam}
show anomalous line ratios in the UV, which can be explained by a
non-solar chemical composition induced by a nova event (Bonnet-Bidaud
\& Mouchet, 1987). Therefore Friedrich et al.~(1996b)
performed an analysis of IUE spectra of \object{V1432\,Aql} but found
no evidence for an anomalous che\-mical composition.

The locking of the white dwarf rotation into synchronization in the
near-synchronous systems may be re-established over time scales of the
order of $10^2$--$10^3$ yr, as expected from the torques caused by the
accretion and the magnetic interaction of the binary components
(Campbell \& Schwope, 1999). Such secular changes in the
spin periods have been detected for the systems \object{V1500\,Cyg},
\object{BY\,Cam}, and \object{V1432\,Aql}, see Table~\ref{nsync} for
the corresponding synchronization time scales $\tau_{\rm sync}$ for the
white dwarfs. From a combined analysis of optical and X-ray data
Geckeler \& Staubert~(1997a) have first detected a marginally
significant (2\,$\sigma$ level) ${\rm dP}_{\rm spin}/{\rm dt}$ of the
white dwarf spin period of the order of $-10^{-8}$\,s/s in
\object{V1432\,Aql}. The corresponding synchronization time scale
$\tau_{\rm sync}$ of 100\,yr is in excellent agreement with the
theoretical prediction from the dominant magnetic torque acting on the
white dwarf. Our new data verify this result and allow a more
accurate determination of ${\rm dP}_{\rm spin}/{\rm dt}$, see
Section~7.

\section{Dipole Accretion Model}

Geckeler \& Staubert~(1997a) have developed a model for the
accretion process in near-synchronous polars, where the infalling
matter is captured in a centered dipole field geometry and accreted
along different field lines during the beat cycle, see Fig.~\ref{pol}.

\begin{figure}
\resizebox{\hsize}{!}{\rotatebox{90}{\includegraphics{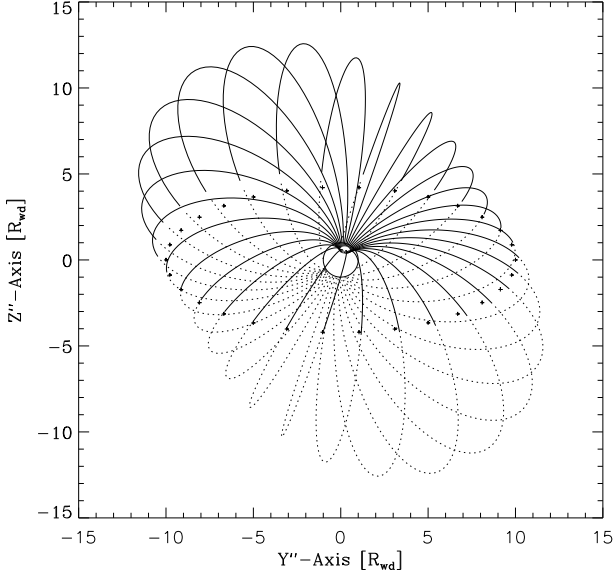}}}
\caption{Accreting field lines in a near-synchronous polar
as seen by an observer in an arbitrary coordinate system X''Y''Z''. 
This corresponds to one particular observing position in the X'Y'Z' 
system with assumed values of $\beta=30^\circ$, an equivalent threading radius
${\rm R}_{\rm t0}'=10\,{\rm R}_{\rm wd}$, and an inclination $\rm{i}=65^\circ$.
Solid field lines are situated above the binary plane, dotted lines below.}
\label{3dpolea}
\end{figure}

\begin{figure}
\resizebox{\hsize}{!}{\rotatebox{90}{\includegraphics{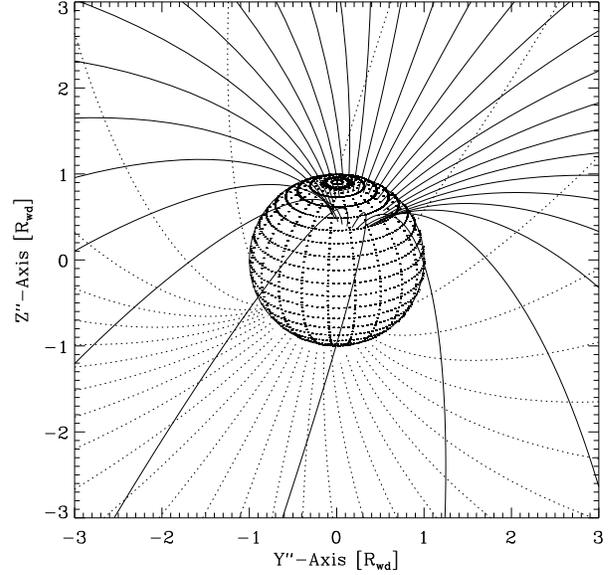}}}
\caption{Accreting field lines in the near-synchronous 
system from Fig.~\ref{3dpolea} with a magnified view on the white dwarf. The 
accretion region on the white dwarf surface traces an ellipsoidal path around 
the axis of the magnetic dipole field during one beat cycle.}
\label{3dpoleb}
\end{figure}

The magnetic moment $\vec{\mu}$ of the white dwarf is inclined
relative to the spin axis (Z$'$-axis) by the colatitude $\beta$ and
positioned in the (X$'<0$) Z$'$ half plane. We have assumed that the
infalling matter (the Horizontal Stream in Fig.~\ref{pol})
is captured in the orbital plane of the binary system at the threading
point which is 
at a distance ${\rm R}_{\rm t}'$ from the white dwarf primary and at
an azimuthal angle $\zeta_{\rm t}'$ relative to the positive X$'$-axis
in the X$'$Y$'$-plane. See Geckeler \& Staubert (1997a) for
further details about the system geometry and the definition of the 
equivalent threading radius ${\rm R}_{\rm t0}'$ which is being used as 
a model parameter. 
For systems with ${\rm P}_{\rm spin} > {\rm P}_{\rm orb}$, and under 
the assumption that the threading point lies on the line of stellar 
centers, the azimuthal angle $\zeta_{\rm t}'$ is 
associated with the beat phase (see the exact definition below)
by $\zeta_{\rm t}'=\Phi_{\rm beat}\cdot2\pi +\pi$. 
(We note that the definition of $\zeta_{\rm t}'$ in 
Geckeler \& Staubert (1997a) is not correct, it should instead read as
here. The definition of beat phase is the same in both papers). 
It is, however, to be
expected that the horizontal stream leaving the inner Lagrangian point
does follow a ballistic trajectory in the orbital plane under some
angle to the line of stellar centers (Lubow \& Shu, 1975, 
Mouchet et al., 1997). In order to be independent of the 
geometric details, we define the beat phase 
$\Phi_{\rm beat}({\rm T})=({\rm T}-{\rm T}_{\rm 0})/{\rm P}_{\rm
  beat}$ through observations: ${\rm T}_{\rm 0}$ (see the ephemeris
given in Table~\ref{fitpar}) is the time when the deepest 
minimum in the optical 
and the X-ray light curves due to spin modulation coincides with the
center of dips (assumed to be eclipses) due to binary modulation. 
As we will discuss below, this means physically that the
observer is looking down the accretion column mostly aligned with the
line of centers of the binary components.
The equivalent radius ${\rm R}_{\rm t0}'$, rather than the
physical radius ${\rm R}_{\rm t}'$, will be used as a model parameter:
\begin{displaymath}
{\rm R}_{\rm t}'={\rm R}_{\rm t0}'
(1+3 \cos^2 (\epsilon_{\rm t}))^{\alpha/2}
\end{displaymath}

During one beat phase the threading point rotates $360^\circ$ in the
orbital plane and the infalling matter is captured progressively by
different field lines. Figure~\ref{3dpolea} shows the accreting field
lines 
as seen by an observer in an arbitrary coordinate system X''Y''Z''. 
This corresponds to one particular observing position in the X'Y'Z' 
system with assumed values of $\beta=30^\circ$, an equivalent
threading radius ${\rm R}_{\rm t0}'=10\,{\rm R}_{\rm wd}$, and an 
inclination $\rm{i}=65^\circ$.
The accretion stream follows different field lines
as a function of beat phase $\Phi_{\rm beat}$ and the impact point
traces an ellipsoidal path around the magnetic pole (center of the
field line bundle) on the white dwarf surface during one beat period
(Fig.~\ref{3dpoleb}). In near-synchronous polars, the accretion rates
for the two poles may vary systematically during the beat cycle and
pole switching may occur.
This effect depends on the location of the threading point in the 
coordinate system of the magnetic dipole field (Cropper, 1989).
Figure~\ref{3dpolea} and Figure~\ref{3dpoleb} show the accreting field lines 
both above and below the orbital plane of the system. 

In addition to the displacement of the accretion region on the white
dwarf surface, the contribution of the inclination of the field line
(and thus the accretion funnel) relative to the surface normal at the
accretion spot has to be taken into consideration, e.g., for cyclotron
emission features in polars and absorption troughs by the funnel.
These features are influenced by the orientation of the magnetic field
in addition to the displacement of the accretion spot. Both effects
result in shifts of the timing of light curve and polarization
features against the spin ephemeris as a function of the beat phase
$\Phi_{\rm beat}$.

\section{Data}

We have obtained a total of 42 nights of optical CCD photometry of 
\object{V1432\,Aql} during the years 1993--1997: 35 nights in white 
light at the T\"ubingen 40\,cm Cassegrain or the 30\,cm refractor 
and 7 nights in R--band at the Sonneberg 60\,cm Cassegrain. 
Standard reduction of the CCD
frames (bias subtraction, dark correction, flat fielding) has been
performed. The CCD frames have been analysed with our own IDL
software package for aperture photometry which automatically finds 
the optimum extraction
radii to maximize the signal-to-noise ratio for point sources in our
mostly background dominated frames. We have achieved time resolutions
ranging from 30\,s to 160\,s. Additionally, photometric data from
Patterson et al.~(1995) and Watson et al.~(1995) have
been scanned and included in the analysis. Timing data have been
barycenter corrected using a software of the Northern Data Reduction
Consortium (NDAC) developed for the Hipparcos project.

\begin{table}
\caption[]{\protect\label{robslog}
Rosat PSPC observations analysed}
\begin{flushleft}
\begin{tabular}{lcccc}\hline
\multicolumn{1}{c}{ROR-ID} &
\multicolumn{1}{c}{Date} &
\multicolumn{1}{c}{JD of obs.$^{\rm b}$} &
\multicolumn{1}{c}{${\rm T}_{\rm int}^{\rm a}$ [ks]} &
\multicolumn{1}{c}{PI}\\ \hline
700782            &92/04&48742.0--42.3&  8.7 &Staubert \\
701090            &92/10&48904.3--26.2& 29.6 &Staubert \\
700923            &93/04&49077.8--79.9& 38.2 &Madejski \\
701460 $^{\rm c}$ &93/10&49274.5--78.9& 31.4 &Staubert \\ \hline
\end{tabular}
\end{flushleft}
a: Source integration time\\
b: JD--2\,400\,000.0\\
c: including 701466, 701472, 701478
\end{table}

With respect to X-ray observations of \object{V1432\,Aql} used for 
this investigation, we give an observation log of the ROSAT PSPC 
data in Table~\ref{robslog}. About 110\,ks of energy-resolved Rosat 
observations with the imaging X-ray instrument PSPC (0.1--2.4\,keV 
energy range) of the polar \object{V1432\,Aql} obtained by us and other
investigators have been used for this paper (see Table~\ref{robslog}). 
The source was also observed with the Rosat/HRI, however, no trough
timings could be determined from these data. In addition,
we make use of one RXTE observation of  12 July 1996 with a 
total on-source time of 21.7~ks. For this observation an off-axis pointing
was chosen in order to avoid contamination by the Active Galaxy 
\object{NGC\,6814} situated 37$'$ away from \object{V1432\,Aql}. 
It places \object{NGC\,6814} at the edge of the field of
view of the PCA instrument (approx. 1$^\circ$ FWHM). 
The light curves cover the energy range
2 to 20\,keV (channels 4--53, epoch 3) and have a temporal resolution of
16\,s (data mode ``standard2f''). The data have been background subtracted
and barycenter corrected. From these data no information about the
spin ephemeris (Section 5.3) have been extracted, they have only been 
used for the period analysis as discussed in Section 5.2. 

\section{Optical and X-ray Spin Modulation}
\subsection{General Results}

\begin{figure}
\resizebox{\hsize}{!}{\rotatebox{90}{\includegraphics{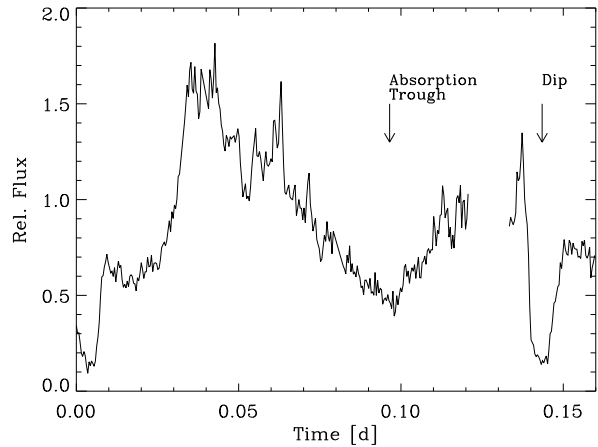}}}
\caption{CCD photometry in white light of \object{V1432\,Aql} 
from 9 Sept. 1996, obtained with the 40 cm Cassegrain at 
T\"ubingen with an integration time of 30\,s.}
\label{lc}
\end{figure}

Figure~\ref{lc} presents a sample optical light curve of
\object{V1432\,Aql} as obtained on 9 Sept. 1996 with the 40 cm Cassegrain 
of the T\"ubingen institute,
demonstrating the complex intensity variations of this object in the
optical waveband.

Narrow 'dips' in the optical and X-ray light curves with full widths
between half intensity at ingress and egress of $\sim700$\,s 
(Geckeler \& Staubert, 1997b, Geckeler \& Staubert, 1999, 
Mukai et al. 2000) and the radial velocities of the narrow emission 
line components (Staubert et al. 1994) follow the orbital period of 
the binary of $\sim12116.3$\,s (Patterson et al., 1995). We take
these 'dips' as eclipses of the emission region by the secondary.

The fast
ingress/egress of the source of the X-ray emission in the RXTE data
during the dip event, the total absorption of the X-ray flux during
dip totality, the non-detection of hardness variations due to
energy-dependent absorption in both RXTE and Rosat data and the stable
timing of the X-ray dips with respect to a linear ephemeris all
support this interpretation. 

The main modulation of the optical and X-ray flux is caused by the
changing aspect of the accretion spot on the surface of the white
dwarf, the emission characteristics of the different emission
mechanisms in the optical and X-ray wavebands, and absorption by the
magnetically confined accretion funnel directly above the accretion
region (see Imamura \& Durisen, 1983 for simulated X-ray pulse
profiles). The broad modulation thus follows the spin period of the
white dwarf ($\sim12150$\,s).

The broad feature in the optical light curve labeled 'trough' with a
full width of 2000--3000\,s between half intensity is assumed to occur
mainly due to absorption in the accretion column, when the viewing
angle between the accretion funnel directly above the accretion region
and the direction to the observer is at a minimum, and due to the
emission characteristics of the optical cyclotron emission.  It is
also prominent in the X-ray waveband, where $\tau\gg 1$ due to
electron scattering and bound-free absorption in the funnel (Imamura
\& Durisen, 1983), see also Geckeler \& Staubert~(1997a)
for a combined optical/X-ray pulse profile of \object{V1432\,Aql}. The
absorption is sufficient to produce the trough feature\ in the X-ray
band for typical physical parameters of the accretion region in
\object{V1432\,Aql}.

We have observed a shift of the trough timings with a
half amplitude of the order of 1000\,s with respect to the spin
ephemeris as a function of the phase $\Phi_{\rm beat}$. 
There are two contributions to the total time shifts: the first
is due to the longitudinal displacement of the hot spot on the white
dwarf surface, and the second is due to the changing orientation of
the field lines guiding the accretion stream.

\subsection{Period Analysis}

\begin{figure}
\resizebox{\hsize}{!}{\includegraphics{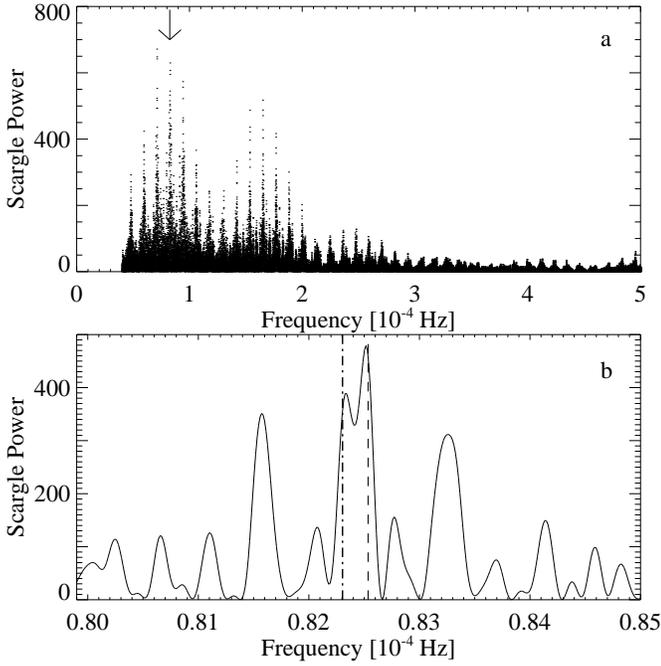}}
\caption{
\textbf{a}
PSD of the optical CCD photometry of V1432\,Aql from 22 nights in 
1996/1997. The arrow marks the 12116\,s orbital period of the system. 
\textbf{b}
Zoom into the PSD under \textbf{a}, centered on the 12\,ks regime, allowing 
to separate orbital and spin periods. The dashed vertical line marks the
12116\,s orbital period, the dash-dotted line indicates the
12150\,s spin period.}
\label{to}
\end{figure}

A periodogram analysis (Scargle, 1982) has been performed on a set of
optical observations of \object{V1432\,Aql} consisting of data from 22
nights spread over 1996--1997 (15 nights in white light in T\"ubingen, 
7 nights in R--band in Sonneberg). For the joint analysis the R--band
data have been transformed into the T\"ubingen ST7-white light system
using overlapping data sets. The resulting periodogram is shown in
Fig.~\ref{to}. Both periods, the orbital period and the spin period, are 
clearly separated.

\begin{figure}
\resizebox{\hsize}{!}{\includegraphics{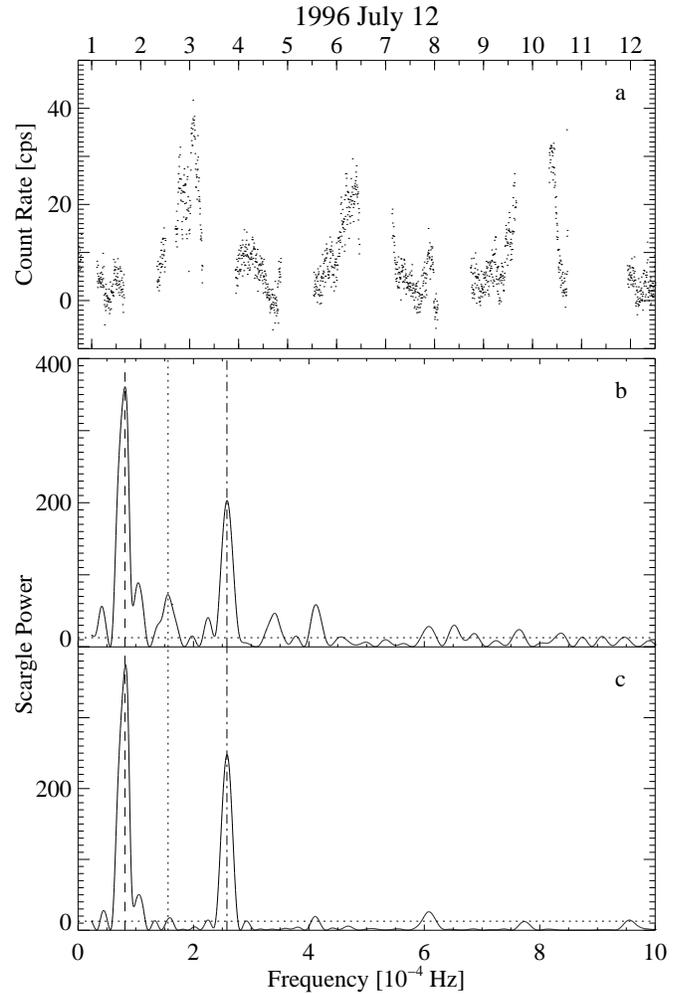}}
\caption{\textbf{a}
Background-subtracted light curve of \mbox{V1432\,Aql}
in the 2--~20\,keV energy range as seen by RXTE. The time axis is labeled
in hours (UT). \textbf{b} PSD of the above observation, calculated
according to Scargle~(1982).
\textbf{c} PSD of a simulated sinusoidal
light curve with a single period of $12150$\,s. In \textbf{b} and
\textbf{c}, the dashed, dotted, and dash-dotted vertical lines indicate
the centers of the peaks in the PSD of the RXTE data
at 12300\,s, 6410\,s and 3880\,s, respectively. The FWHMs of the peaks are
3800\,s, 1070\,s and 350\,s. The dotted horizontal
line corresponds to a significance of $99\%$ for a periodic signal.
For further details see text.}
\label{rxtelc}
\end{figure}

Timing analyses of the X-ray flux of \mbox{V1432\,Aql} have been
presented, e.g., by Staubert et al.~(1994), Friedrich et
al.~(1996a) and Mukai~(1998). In doing such an analysis, it is quite
important to use the method of Scargle~(1982) for unevenly sampled
data. Since the
periods of interest are in the $12$\,ks range, with the
satellite orbit being $\sim5.8$\,ks, the sampling considerably
influences the timing analysis.  This has also been the case with
earlier Rosat observations analysed by Mukai~(1998).

In the following analysis we present X-ray data with high signal to
noise ratio from one of our RXTE observations which was done on 
12 July 1996 with a total on-source time of 21.7~ks
(we concentrate on this particular observation since it
is least affected by observational gaps).

Figure~\ref{rxtelc}a shows the light curve in the 2--20 keV range and 
Figure~\ref{rxtelc}b the corresponding Scargle Power Spectral 
Density (PSD). In order to make 
comparisons easier, we display frequency instead of period dependence 
and show the same frequency interval as Mukai, 1998. 

The PSD contains several peaks well above the PSD level of $\sim13$,
corresponding to a $1\%$ false alarm probability (fap), i.e., a
significance of $99\%$ for a periodic signal in the data. 
The fap was computed using the statistical distribution of the peak 
Scargle power of $1000$ simulated realizations of a Gaussian white 
noise process with the same variance and the same sampling as the 
original light curve.
The peaks in the PSD are asymmetric and very broad due to the window
function. The most prominent peak is centered at $\sim12300$\,s with a
FWHM of 3800\,s. As can be expected, it is not possible to separate
the $12116$\,s orbital period and the $12150$\,s period. The same has
been the case for the Rosat data used by Mukai~(1998), which had led
to the proposal of \object{V1432\,Aql} being an intermediate polar (IP) 
with ${\rm P}_{\rm spin}\approx 4040-4050$\,s. A strong peak
corresponding to a period of $\sim4$\,ks is apparent also in
the RXTE data, but it reaches only $56\%$ of the peak power of the
12\,ks feature in the PSD.  The local maximum of the PSD near 4\,ks is
located at 3880\,s, with 350\,s FWHM of the peak. The power in the
6\,ks regime is comparable to that of several smaller peaks introduced
by the window function. This regime coincides with the satellite orbit
of RXTE ($\sim5.8$\,ks), as has also been the case with Rosat.

In order to test whether the 4\,ks period is a true period,
we calculated the PSD of simulated data assuming a sinusoidal 
light curve with a period of $12150$\,s and the same sampling 
and signal-to-noise ratio as the RXTE observation of \object{V1432\,Aql}. 
It is clear from this PSD, given in Fig.~\ref{rxtelc}c, that even 
with the simple assumption of a sine modulation with a single 
period of $12150$\,s, the 4\,ks feature shows up as an alias 
of the longer period: the simulation reproduces the height and width 
of the peak, as well as its location in the frequency domain. 
We therefore conclude that the 4\,ks feature is most likely not a true 
period. In addition to these arguments from the period analysis 
of our optical and X-ray observations, we will 
show in Section~7 that the synchronization time scale observed 
in \object{V1432\,Aql} also strongly argues against the interpretation 
of the system as an Intermediate Polar.

\subsection{Spin Ephemerides}

In order to analyse changes in the shape and timing of the trough as a
function of the varying accretion geometry in \object{V1432\,Aql}, we
have defined procedures to get accurate estimates of the time of
trough minimum and its width. We have defined T$_{\rm hi}$ as the
time of half intensity between the flux at the onset of the trough and
the flux at the bottom of the trough, and, equivalently, T$_{\rm he}$ 
as the time of half
intensity during the phase of recovering flux. The full width of the
trough profile at half intensities is given by $\Delta {\rm
  t}_{\rm fwhie}={\rm T}_{\rm he}-{\rm T}_{\rm hi}$.

\begin{table}
\caption[]{\protect\label{ecl}
Optical and X-ray trough timings}
\begin{flushleft}
\begin{tabular}{lrr} \hline
\multicolumn{1}{c}{${\rm T}_{\rm bs}^{\rm a}$} &
\multicolumn{1}{c}{Error ${\rm T}^{\rm b}_{\rm bs}$ [s]} &
\multicolumn{1}{c}{$\Delta {\rm t}^{\rm fwhie}$ [s]} \\ \hline
\multicolumn{3}{c}{White Light CCD Photometry T\"ubingen} \\ \hline
49543.54387  &  600  &  -  \\
49567.45594  &  600  &  -  \\
49568.43930  &  400  &  -  \\
49638.32219  &  600  &  -  \\
50281.40729  &   140  &  2200  \\
50285.48114  &   170  &  -     \\
50286.46490  &   160  &  3400  \\
50287.44850  &   120  &  3200  \\
50332.30930  &   170  &  2200  \\
50332.45138  &    90  &  3000  \\
50336.38632  &   220  &  2900  \\
50639.56572  &   200  &  -     \\ \hline
\multicolumn{3}{c}{R-Band CCD Photometry Sonneberg} \\ \hline
50285.47960  &  120  &  3000  \\
50286.46433  &  150  &  3100  \\
50287.44977  &  200  &  2900  \\ \hline
\multicolumn{3}{c}{V-Band CCD Photometry,
Patterson et al.~(1995)} \\ \hline
49571.67324  &   270  &  4200  \\ 
49572.65849  &   150  &    -   \\ 
49577.71404  &   130  &  2900  \\ 
49577.85521  &   170  &  2700  \\ 
49578.69991  &   140  &  2900  \\ 
49578.83958  &   200  &  2900  \\ 
49579.68234  &   200  &  2700  \\ 
49579.82152  &   190  &  2700  \\ 
49580.66430  &   150  &  2800  \\ 
49580.80703  &   140  &  3200  \\ 
49581.65049  &   140  &  3300  \\ 
49581.78985  &   140  &  2800  \\ 
49582.77559  &   210  &  3100  \\ \hline
\multicolumn{3}{c}{White Light Photometry,
Watson et al.~(1995)} \\ \hline
49238.66143  &   400  &     -  \\
49239.50662  &   340  &     -  \\ \hline
\multicolumn{3}{c}{Rosat Data} \\ \hline
48921.52856 & 180 & 3200 \\
49078.74851 & 110 & 2200 \\
49277.04638 & 170 & 2700 \\ \hline
\end{tabular}
\end{flushleft}
a: BJD--2\,400\,000\\
b: Error estimates, see text\\
\end{table}

\begin{figure}[htb]
\resizebox{\hsize}{!}{\rotatebox{0}{\includegraphics{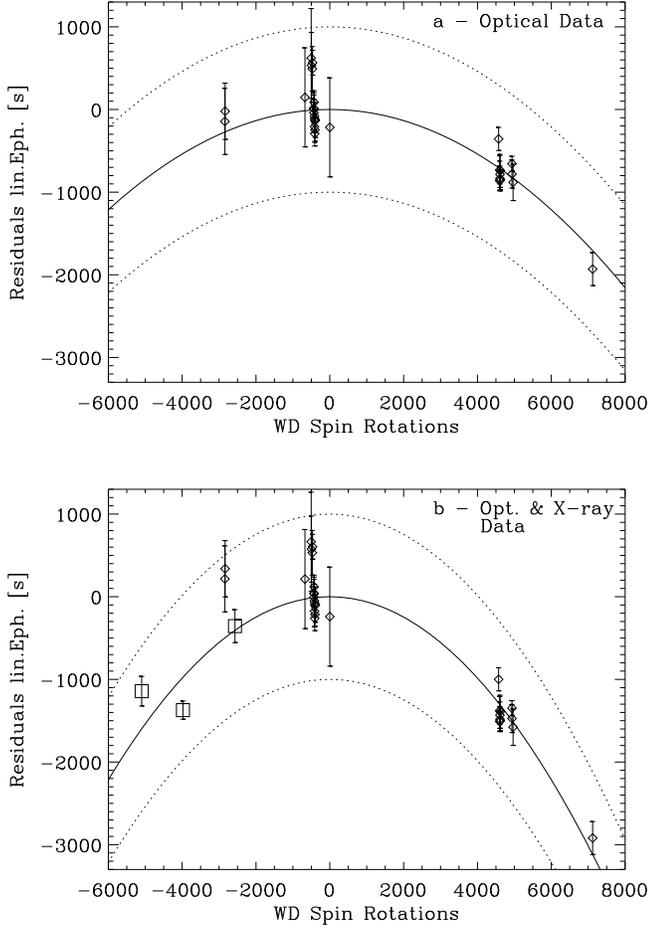}}}
\caption{\protect
\textbf{a} Residuals of the timings ${\rm T}_{\rm bs}$ of trough
minima in the
optical photometry with respect to the linear component
of the quadratic best fit ephemeris (ephemeris (1) in Table~\ref{spin}).
The solid line represents the quadratic ephemeris, the dotted
lines mark the maximum deviation of the trough timings
${\rm T}_{\rm bs}$ as caused by the varying accretion geometry.
\textbf{b}
Residuals of the timings ${\rm T}_{\rm bs}$ of trough minima in the
optical photometry and X-ray data (square symbols) with respect
to the linear component of the best fit ephemeris (3)
in Table~\ref{spin}.}
\label{aspinres}
\end{figure}

\begin{table}[hbt]
\caption[]{\label{spin}
Ephemerides for the trough minima ${\rm T}_{\rm bs}$}
\begin{flushleft}
\begin{tabular}{lllr} \hline
\multicolumn{1}{c}{${\rm P}_{\rm spin}$ [d]} &
\multicolumn{1}{c}{$\dot{\rm P}_{\rm spin} [10^{-8}$\,s/s]} &
\multicolumn{1}{c}{${\rm E}_{\rm 0}^{\rm a}$} &
\multicolumn{1}{c}{Data}\\ \hline
0.14062688(90) & --0.55(28)  & 49638.32469(98)          & (1)\\
0.14062958(94) & --0.98(29)  & 49549.7361(41)           & (2)\\
0.14062845(26) & --1.013(98) & 49638.32497(93)          & (3)\\ \hline
\end{tabular}
\end{flushleft}
The 1$\sigma$ errors of the parameters are 
given in terms of the last two significant
digits (in parentheses).\\
a: BJD-2\,400\,000.\\
(1) Optical trough timings in CCD photometry from T\"ubingen and
Sonneberg (1993--97) and the scanned photometry from
Patterson et al.~(1995) and Watson et al.~(1995);
(2) Geckeler \& Staubert~(1997a), combined fit to a subset of the 
optical trough timings and X-ray trough timings;
(3) Combined fit to all available optical trough timings
from (1) plus the X-ray trough timings.
\end{table}

The timing of the trough minimum T$_{\rm bs}$ has been defined by the
bisector method to provide an accurate estimate of the minimum. The
trough profile is successively cut at different intensity levels and
the mean timings between identical flux levels at decreasing and
increasing flux are determined, which define the bisector line. The
timing T$_{\rm bs}$ is then defined as the intersection of a straight
line fit through the bisector with the flux curve during the trough
minimum. This procedure refines the determination of the minimum
timing when compared to a subjective estimate of the trough center.
Errors of the trough minima T$_{\rm bs}$ have been estimated according
to the quality and time resolution of the light curves. They are of the
order of 5\% of the full widths $\Delta {\rm t}_{\rm fwhie}$ of the
troughs.

Table~\ref{ecl} gives the resulting timings T$_{\rm bs}$ and full
widths $\Delta {\rm t}_{\rm fwhie}$ for the troughs in the CCD
photometry from T\"ubingen and Sonneberg from 1993--97 plus the
timings derived from the photometry of Patterson et
al.~(1995) and Watson et al.~(1995). The Rosat PSPC
profiles have been obtained by folding flux data in the 0.5--2.4\,keV
energy range with the spin period to obtain pulse profiles with a
resolution of 50 phase bins.  Due to gaps in our RXTE observations we 
were not confident in trough timings from these data.
The application of the bisector method to define the timings T$_{\rm
  bs}$ of the trough minima improved the quality of the timings
compared to the values used in Geckeler \& Staubert~(1997a) 
(note that in Table~1 of that paper entries 1, 4 and 7 are not valid). 

The timings $T_{\rm bs}$ of the troughs from Table~\ref{ecl} have 
been analysed by fitting a quadratic spin ephemeris to the data:
\begin{displaymath}
{\rm T}_{\rm bs}={\rm E}_{\rm 0} + {\rm N}\cdot {\rm P_{\rm spin}} +
{\rm N}^2\cdot\left( \frac{{\rm P_{\rm spin}}\dot {\rm P}_{\rm spin}}{2}\right)
\end{displaymath}

The resulting ephemerides are given in Table~\ref{spin}. There is
clear evidence for a secular change $\dot {\rm P}_{\rm spin}$ of the
white dwarf spin period of the order of $-10^{-8}$\,s/s, with a
resulting synchronization time scale of $10^2$\,yr, comparable to the
170\,yr seen in \object{V1500\,Cyg}.

Figure~\ref{aspinres}a shows the residuals of the timings ${\rm
  T}_{\rm bs}$ of the trough minima from the optical photometry with
respect to the linear component of the quadratic best fit ephemeris
(1) from Table~\ref{spin}. The solid line represents the quadratic
ephemeris, the dotted lines mark the maximum deviations of the trough
timings ${\rm T}_{\rm bs}$ as caused by the varying accretion
geometry. Figure~\ref{aspinres}b shows the residuals of the timings
from the combined optical and the X-ray data set with respect to the
linear component of the best fit ephemeris (3) from Table~\ref{spin}.

\section{Accretion Geometry in \object{V1432\,Aql}}

In this section, we will apply the dipole accretion model plus a
quadratic spin ephemeris to the timings ${\rm T}_{\rm bs}$ of the
trough minima from Table~\ref{ecl} by means of a simultaneous best fit
procedure, starting with the analysis of the optical trough data.

\begin{figure}[htb]
\resizebox{\hsize}{!}{\includegraphics{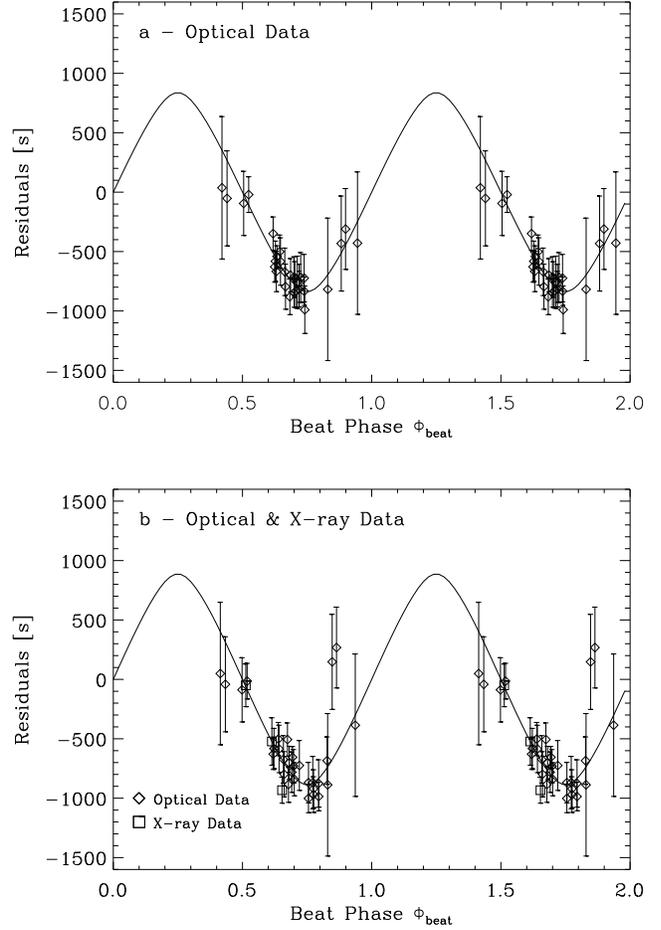}}
\caption{\textbf{a}
  Residuals of the timings ${\rm T}_{\rm bs}$ of the optical trough
  minima with respect to the quadratic spin ephemeris as a function of
  the beat phase $\Phi_{\rm beat}$ (parameters and phase according to
  model MO from Table~\ref{fitpar}). The solid line shows the best fit
  dipole accretion model.  \textbf{b} Results for the combined fit of
  the model to the optical and X-ray trough data (parameters and phase
  according to model MXO from Table~\ref{fitpar}).}
\label{model}
\end{figure}

Figure~\ref{model}a shows the residuals of the estimated timings ${\rm
  T}_{\rm bs}$ of the optical trough minima with respect to the
quadratic spin ephemeris as a function of the beat phase $\Phi_{\rm
  beat}$ (solid line: dipole accretion model for the best fit
parameters MO in Table~\ref{fitpar}). The $\chi_{\rm red}^2=0.34$ for
the fit (24 degrees of freedom = 30 data points minus 6 free
parameters) clearly shows that the model adequately describes the
trough timings ${\rm T}_{\rm bs}$ over the whole five year period
1993--1997. However, we have overestimated the uncertainties of the
trough timings, which leads to $\chi_{\rm red}^2<1$.

The assumption of a dipole field for the white dwarf field geometry in
\object{V1432\,Aql} is appropriate to describe the shifts of the
trough timings, but due to the limited coverage of the beat phase
$\Phi_{\rm beat}$ it is not possible to rule out more complex field
geometries.  Due to the limited beat phase coverage, we also cannot
rule out the presence of pole switching in the system.  The accretion
rates for the two poles are supposed to vary systematically during the
beat cycle, or the accretion even may switch poles altogether at
selected beat phases, depending on the hemispheric location of the
threading point in the magnetic coordinate system of the dipole field.
The model in Fig.~\ref{model} has been calculated for the magnetic
pole facing the threading region at $\Phi_{\rm beat}=0.0$ 
(see Fig.~\ref{pol} for a visualization). The bulk of trough timings 
are located between beat
phase 0.4 and 0.9, where a simple pole switching model would assume
this pole to be the main accreting pole. Clearly, a more complete
coverage of $\Phi_{\rm beat}$ is necessary to search for the effects
of pole switching.

According to this simplified pole-switching scenario, almost all
trough timings observed in \object{V1432\,Aql} are thus associated
with the same accreting pole. We therefore conclude that we mainly
observed the effects of the varying accretion geometry during the beat
cycle for one pole. For an inclination ${\rm i}\neq 90^\circ$, one
pole will be predominantly visible throughout the spin cycle. For
this pole, the observer will look directly into the accretion funnel
near beat phase $\Phi_{\rm beat}=0.0$ for geometries with ${\rm
  i}\approx\beta+1.5\cdot\delta$. The angle $\delta$ is defined by the
equation $\sin{\delta}=({\rm R}_{\rm t}'/{\rm R}_{\rm
  wd})^{-1/2}\cos{\beta}$. It is the angle between the dipole axis
and the actual location of the accretion spot at $\Phi_{\rm
  beat}=0.0$. The factor 1.5 results from the inclination of the
accretion funnel relative to the surface normal, which is
approximately $\delta/2$ for a dipole field geometry. An absorption
of the emission from the accretion region near this pole by the funnel
(assumed to cause the trough features in our model) is
possible for observing geometries with ${\rm i}\geq
\beta+1.5\cdot\delta$. For estimating $\delta$ from such observations 
more data are needed.

\begin{table}[hbt]
\caption[]{\label{fitpar}
Best fit parameters for the dipole accretion model}
\begin{center}
\begin{tabular}{cccc}\hline
\multicolumn{4}{c}{Optical Trough Timings -- Parameter Set MO}\\ \hline
Parameter & Best Fit & $-1$\,$\sigma$ & +1\,$\sigma$ \\ \hline
${\rm R}_{\rm t0}'\,\,[{\rm R}_{\rm wd}]$ &
12.6 & 6.9 & - \\
$\beta$\,\,[$^\circ$] &
85.1 & - & - \\
$\sqrt{{\rm R}_{\rm t0}'/{\rm R}_{\rm wd}}\cdot\sin{\beta}$ &
3.6 & 1.1 & 2.7 \\
${\rm P}_{\rm spin}$\,\,[d] &
0.1406271 & $1.6\cdot10^{-6}$ & $1.2\cdot10^{-6}$ \\
$\dot {\rm P}_{\rm spin}$\,\,[s/s] &
$-5.4\cdot10^{-9}$ & $3.2\cdot10^{-9}$ & $3.7\cdot10^{-9}$ \\
${\rm E}_{\rm 0spin}$\,[BJD] &
2\,449\,638.3317 & $6.2\cdot10^{-3}$ & $7.3\cdot10^{-3}$ \\
${\rm P}_{\rm beat}$\,[d] &
50.28 & 0.15 & 0.21 \\
${\rm E}_{\rm 0beat}$\,[BJD] &
2\,449\,646.9 & 4.3 & 6.2 \\ \hline
\multicolumn{4}{c}{Optical and X-ray Trough Timings -- MXO}\\ \hline
${\rm R}_{\rm t0}'\,\,[{\rm R}_{\rm wd}]$ &
11.2 & 5.0 & - \\
$\beta$\,\,[$^\circ$] &
89.2 & - & - \\
$\sqrt{{\rm R}_{\rm t0}'/{\rm R}_{\rm wd}}\cdot\sin{\beta}$ &
3.5 & 0.9 & 1.9 \\
${\rm P}_{\rm spin}$\,\,[d] &
0.14062879 & $5.0\cdot10^{-7}$ & $5.2\cdot10^{-7}$ \\
$\dot {\rm P}_{\rm spin}$\,\,[s/s] &
$-1.01\cdot10^{-8}$ & $1.5\cdot10^{-9}$ & $1.5\cdot10^{-9}$ \\
${\rm E}_{\rm 0spin}$\,[BJD] &
2\,449\,638.3325 & $4.4\cdot10^{-3}$ & $8.8\cdot10^{-3}$ \\
${\rm P}_{\rm beat}$\,[d] &
50.062 & 0.066 & 0.064 \\
${\rm E}_{\rm 0beat}$\,[BJD] &
2\,449\,646.9 & 5.9 & 3.0 \\ \hline
\end{tabular}
\end{center}
\end{table}

\begin{figure}[htb]
\resizebox{\hsize}{!}{\includegraphics{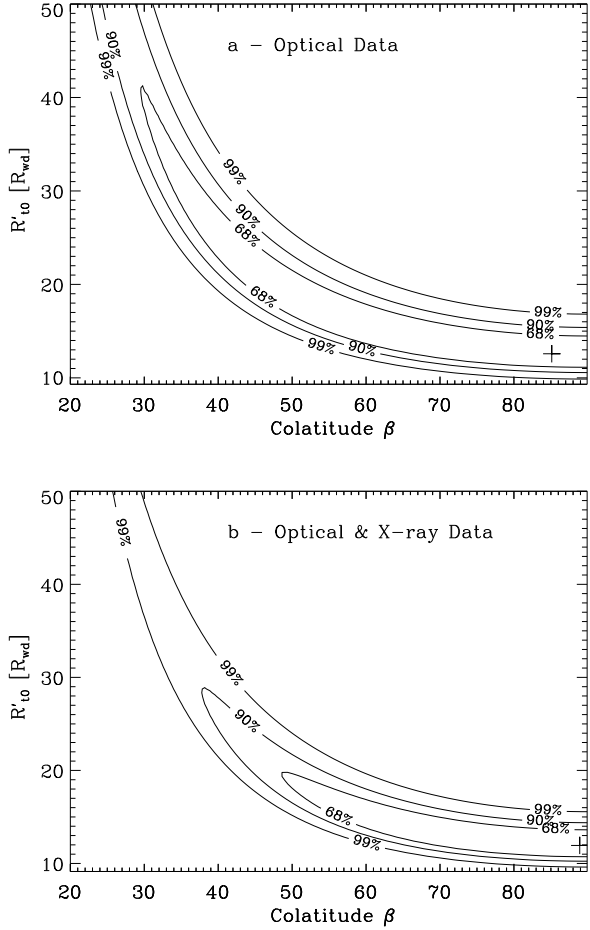}}
\caption{
\textbf{a}
Isocontours of the $\chi^2$ distribution
as a function of the model parameters $\beta$ (in degrees) and
${\rm R}_{\rm t0}'$ for the description of the optical
trough timings by the dipole accretion model plus a
quadratic ephemeris. The contours clearly demonstrate the
$({\rm R}_{\rm t0}'/{\rm R}_{\rm wd})^{1/2}\sin{\beta}$ relation
between the two model parameters. The contour lines have been calculated
for confidence levels of 68\%, 90\% and 99\%.
The absolute minimum of the $\chi^2$ has been marked by a cross.
\textbf{b}
Isocontours of the $\chi^2$ distribution for
the optical and X-ray trough timings combined.}
\label{chisq}
\end{figure}

The resulting best fit parameters for the simultaneous application of
the dipole accretion model plus a quadratic ephemeris to the optical
trough timings are given in Table~\ref{fitpar}, together with their
jointly estimated $1\,\sigma$ uncertainties ($\chi^2_{\rm min}$+7.01
for six free parameters). The optical data allow us to determine the
spin period of the white dwarf ${\rm P}_{\rm spin}=12150.18\pm0.14$\,s
and the resulting beat period is ${\rm P}_{\rm beat}=50.28\pm0.21$\,d
(both valid for Epoch 1995.2). We have used ${\rm P}_{\rm
  orb}=0.1402349\pm10^{-7}$\,d (12116.295$\pm0.009$\,s) 
and the reference center of eclipse ${\rm E}_{\rm 0}=$ 
HJD 2449568.509764(1) from Patterson et al.~(1995). 
The ephemeris for the optical trough data
includes a significant secular change of the white dwarf spin period
of ${\rm dP}_{\rm spin}/{\rm dt}=-5.4^{+3.7}_{-3.2}\cdot10^{-9}$\,s/s.

The parameters ${\rm R}_{\rm t0}'$ and $\beta$ are not well
constrained individually by the fit, which is reflected by the large
or not defined $\pm 1\,\sigma$ errors in Table~\ref{fitpar}, because
they are related by the following, purely geometrical expression:

\begin{eqnarray}
\sqrt{\frac{{\rm R}_{\rm t0}'}{{\rm R}_{\rm wd}}}\cdot\sin{\beta} \approx
{\rm const.}
\end{eqnarray}

For a given amplitude of the observed shift of the trough timings, an
increase in the colatitude $\beta$ can thus be compensated for by an
increase in the size of the ellipsoid traced by the accretion region
on the white dwarf surface around the magnetic pole (and thus a
decrease in ${\rm R}_{\rm t0}'$).

The influence of the above relation especially affects the
determination of the colatitude $\beta$, which cannot be constrained
by the fitting procedure. The graph of the shifts of the trough
timings resulting from the dipole accretion model retains a sinusoidal
shape for a wide range of colatitudes. Its amplitude is increasing as
the ellipse traced by the accretion spot (Fig.~\ref{3dpoleb}) is
nearing the rotational axis of the white dwarf with decreasing
colatitude $\beta$. A marked asymmetry in the graph of the shifts of
the trough timings results only if the ellipse traced by the accretion
region is located near the rotational axis ($\beta\approx\beta_{\rm
  crit}$). The critical colatitude $\beta_{\rm crit}$ is defined as
$\tan{\beta_{\rm crit}}=({\rm R}_{\rm t}'/{\rm R}_{\rm wd})^{-1/2}$,
separating accretion geometries in which the white dwarf spin axis is
located outside of the accretion ellipse ($\beta > \beta_{\rm crit}$)
or inside of it ($\beta < \beta_{\rm crit}$). The fitting procedure
is not able to produce meaningful constraints for the colatitude, 
it gives a formal value
$\beta\approx90^\circ$ (which cannot be taken at face value)
unconstraining $\pm 1\,\sigma$ errors. In interpreting the mean period 
between successive meridian crossings of the accretion region one
needs to distinguish between the two cases ($\beta > or < 
\beta_{\rm crit}$) as discussed in Geckeler \& Staubert (1997a).

Because the mean period of the trough timings in \object{V1432\,Aql}
is longer than the orbital period, we obtain $\beta>\beta_{\rm crit}$.
Additionally, an upper $1\,\sigma$ limit (68\% confidence) of
$\beta_{\rm crit}<23^\circ$ for the colatitude can be obtained from
the analysis of the data.

To demonstrate the effect of the $({\rm R}_{\rm t0}'/{\rm R}_{\rm
  wd})^{1/2}\sin{\beta}$ relationship with respect to the optical data
set, the resulting elongated valley in $\chi^2$ as a function of the
model parameters ${\rm R}_{\rm t0}'$ and $\beta$ is illustrated in
Fig.~\ref{chisq}a. The contour lines are for 68\%, 90\% and 99\%
confidence levels ($\chi^2_{\rm min}$+2.28, 4.61 and 9.21 for two free
parameters). The $\chi^2_{\rm min}$ position is marked by a cross.
To emphasize the $({\rm R}_{\rm t0}'/{\rm R}_{\rm
  wd})^{1/2}\sin{\beta}$ relation, the other parameters of the model
have been fixed to their best fit values. This $\chi^2$ valley is
narrower compared to the one resulting from a fit, where all six
parameters are set free (as used in the determination of the best fit
parameters).

\begin{figure}[htb]
\resizebox{\hsize}{!}{\rotatebox{90}{\includegraphics{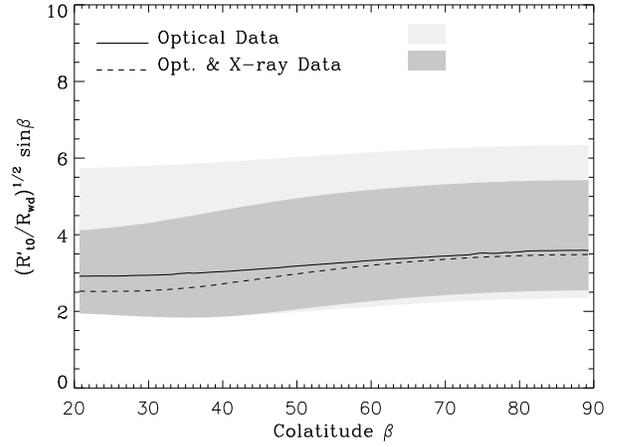}}}
\caption{Parameter
$({\rm R}_{\rm t0}'/{\rm R}_{\rm wd})^{1/2}\sin{\beta}$ 
as a (weak) function of the colatitude $\beta$ (in degrees) as 
derived from a fit
of the dipole accretion model plus a quadratic ephemeris to
the trough timings in the optical data set and a fit to the
combined optical and X-ray data, see text. The 1$\,\sigma$ error
contours (68\% confidence probability) have been marked as grey
areas (partially overlapping).}
\label{cvec}
\end{figure}

The fitting procedure nevertheless does allow to determine the
dimensionless parameter $({\rm R}_{\rm t0}'/{\rm R}_{\rm
  wd})^{1/2}\sin{\beta}=3.6_{-1.1}^{+2.7}$ for the optical data with
the constraint $\beta>\beta_{\rm crit}$. Its value determines the
amplitude of the shift of the trough timings ${\rm T}_{\rm bs}$ in the
dipole accretion model.

The relation between the parameters ${\rm R}_{\rm t0}'$ and $\beta$ is
an approximation for $\beta > \beta_{\rm crit}$, with the value $({\rm
  R}_{\rm t0}'/{\rm R}_{\rm wd})^{1/2}\sin{\beta}$ being a weak
function of $\beta$ (or ${\rm R}_{\rm t0}'$ alternatively).
Figure~\ref{cvec} shows the resulting best fit values for the
parameter $({\rm R}_{\rm t0}'/{\rm R}_{\rm wd})^{1/2}\sin{\beta}$ as a
function of the colatitude $\beta$ for the application of the dipole
accretion model plus a quadratic ephemeris to the optical trough
timings and to the combined optical/X-ray data set. The grey areas
visualize the $\pm1\,\sigma$ error ranges (68\% confidence contours).
As the graph shows, the $({\rm R}_{\rm t0}'/{\rm R}_{\rm
  wd})^{1/2}\sin{\beta}$ values from the analysis of the optical data
set and of the combined optical/X-ray data are consistent. This
dimensionless parameter is the relevant parameter constraining the
accretion geometry of \object{V1432\,Aql}. Because it is, as a first
order approximation, only a function of the observed amplitude of the
shift of the trough timings ${\rm T}_{\rm bs}$ with respect to the
spin ephemeris, its value is constrained by the data in a robust way.
With an independent estimate for $\beta$ (e.g. from polarimetry),
${\rm R}_{\rm t0}'$ could be calculated from this parameter.

The fit also results in ${\rm R}_{\rm t0}'>10\,{\rm R}_{\rm wd}$ at the
68\% confidence level. Together with an estimate of ${\rm R}_{\rm
  t0}'$ as a function of the system parameters from
Mukai~(1988) and of the accretion rate from the ${\rm P}_{\rm
  orb}$--$\dot {\rm M}$ relation ($\dot {\rm M}\propto {\rm P}_{\rm
  orb}^{3.2}$) from Patterson~(1984) of $\dot {\rm M}=2\cdot
10^{16}$\,g s$^{-1}$, this results in ${\rm B}_{\rm pol}> 
4~10^7$\,G for the polar strength of the white dwarf magnetic 
field, as expected for a polar.

Since only three trough timings are available from
Rosat X-ray observations, we have combined them in the analysis with
the optical data set, assuming that the optical and X-ray troughs
occur at exactly the same time.  This assumption may not be quite correct,
as the detailed discussion in Section~7 will show.  We have
applied the dipole accretion model plus a quadratic ephemeris to the
timings ${\rm T}_{\rm bs}$ of the trough minima from Table~\ref{ecl}.
The $\chi_{\rm red}^2=1.1$ (at 27 degrees of freedom = 33 data points
minus 6 free parameters) shows that the dipole accretion model
adequately describes the data set, albeit with a worse fit between the
model and the data compared to the fit to the coherent optical data
set alone.

Figure~\ref{model}b shows the residuals of
the timings ${\rm T}_{\rm bs}$ of the optical and X-ray trough
minima with respect to the quadratic ephemeris as a function of
the beat phase $\Phi_{\rm beat}$ (model parameters MXO,
see Table~\ref{fitpar}).

The resulting best fit parameters are given in
Table~\ref{fitpar}, together with their $1\,\sigma$ errors.  The
combined analysis of the optical and X-ray data allows to determine
the spin period of the white dwarf as ${\rm P}_{\rm
  spin}=12150.328\pm0.045$\,s (Epoch 1995.2), consistent with the
result from the optical data set within error limits. The ephemeris
includes a significant ${\rm dP}_{\rm spin}/{\rm dt}=(-1.01\pm
0.15)\cdot10^{-8}$\,s/s. The absolute value for the period change is
larger than the one derived from the optical data set alone, which
covers a smaller time base. We will discuss this discrepancy in the
estimates of ${\rm dP}_{\rm spin}/{\rm dt}$ in the following Section~7
in detail.

The parameters ${\rm R}_{\rm t0}'$ and $\beta$ are not
constrained individually by the fitting procedure, but the
dimensionless parameter $({\rm R}_{\rm t0}'/{\rm R}_{\rm
  wd})^{1/2}\sin{\beta}=3.5_{-0.9}^{+1.9}$ can be determined,
consistent with the result from the optical data set within error
limits.

The effect of the $({\rm R}_{\rm t0}'/{\rm R}_{\rm
  wd})^{1/2}\sin{\beta}$ relationship is  demonstrated in
Fig.~\ref{chisq}b. The contour lines are for 68\%, 90\% and 99\%
confidence levels and the remaining parameters of the model have been
fixed to their best fit values. The $\chi^2_{\rm min}$ position is
marked by a cross.

\section{Synchronization Time Scale and Torques}

Geckeler \& Staubert~(1997a) provided first evidence
for a secular change of the spin period of the white dwarf
${\rm dP}_{\rm spin}/{\rm dt}=(-0.98\pm 0.29)\cdot10^{-8}$\,s/s
in the system \object{V1432\,Aql}, corresponding to a synchronization
time scale
$\tau_{\rm sync}=({\rm P}_{\rm spin}-
{\rm P}_{\rm orb})/\dot {\rm P}_{\rm spin}$ for an adjustment
of the spin period to the orbital period of
the order of 100\,yr.

Table~\ref{tau} shows the values for ${\rm dP}_{\rm spin}/{\rm dt}$
and the corresponding synchronization time scales
$\tau_{\rm sync}$ derived from the timings ${\rm
T}_{\rm bs}$ of the troughs in the optical data set and a combined
analysis of trough timings in the optical and X-ray data. The
data have either been analysed by fitting a quadratic ephemeris to the
timings ${\rm T}_{\rm bs}$ or the simultaneous application of the
dipole accretion model for a refined fit.

\begin{table}[hbt]
\caption[]{\label{tau}
${\rm dP}_{\rm spin}/{\rm dt}$ and synchronization time scales
$\tau_{\rm sync}$}
\begin{center}
\begin{tabular}{rclc} \hline
\multicolumn{1}{c}{${\rm dP}_{\rm spin}/{\rm dt}^{\rm a}$} &
\multicolumn{1}{c}{\,\,\,\,} &
\multicolumn{1}{c}{$\tau_{\rm sync}$ [yr]} &
\multicolumn{1}{c}{Ref.} \\ \hline
\multicolumn{4}{c}{Optical Data Set} \\ \hline
$-0.55\pm 0.28$ & &
$194^{+176}_{-65}$ & (1) \\
$-0.54\pm 0.37$ & &
$199^{+441}_{-75}$ & (2) \\ \hline
\multicolumn{4}{c}{Optical \& X-ray Data Combined} \\ \hline
$-0.98\pm 0.29$ & &
$110^{+42}_{-25}$ & (3) \\
$-1.013\pm 0.098$ & &
$106^{+11}_{-9}$ & (4) \\
$-1.01\pm 0.15$ & &
$107^{+19}_{-14}$ & (5) \\ \hline
\multicolumn{4}{c}{Optical \& X-ray data (shift included)} \\ \hline
$-0.70\pm 0.28$ & &
$153^{+102}_{-44}$ & (6) \\ \hline
\end{tabular}
\end{center}
a: In units of $10^{-8}$\,s/s\\
(1) All optical data 1993--1997, quadratic ephemeris;
(2) Data set (1), simultaneous application of dipole accretion model;
(3) Geckeler \& Staubert~(1997a),
CCD photometry T\"ubingen 1993--1996, including scanned photometry
from Patterson et al.~(1995) and Watson et al.~(1995)
and the X-ray data from Rosat, dipole accretion model;
(4) All optical data from 1993--1997 and X-ray data from Rosat,
quadratic ephemeris;
(5) Data set (4), simultaneous application of dipole accretion model;
(6) Data set (4), simultaneous application of dipole accretion model
and a shift between the optical and X-ray trough timings. See text
for further details.
\end{table}

Due to the extended data base for this analysis as compared to the data
available to Geckeler \& Staubert~(1997a), a significant
secular change of the white dwarf spin period can now be inferred from
the optical trough timings alone, without using the X-ray data set.
The inclusion of the trough timings in the X-ray waveband expands the
available timebase (X-ray data cover the period 1992.8--93.2) for the
determination of ${\rm dP}_{\rm spin}/{\rm dt}$, resulting in a lower
formal error for the period change. However, the combination of the
X-ray data with the optical data set consistently yields lower values
for the synchronization time scale $\tau_{\rm sync}$ (larger values for
$\mid {\rm dP}_{\rm spin}/{\rm dt}\mid$) than fits to the optical data
only. To examine this discrepancy, we have expanded the dipole
accretion model by allowing for systematic shifts of the X-ray trough
timings ${\rm T}_{\rm bs}$ with respect to the optical data.
A shift of the X-ray trough timings with respect to the optical
troughs of $-700\pm 580$\,s is marginally significant ($\chi^2_{\rm
  red}=0.61$ and $\chi^2=15.8$ with the shift, $\chi^2_{\rm red}=1.12$
and $\chi^2=30.2$ without). The resulting ${\rm dP}_{\rm spin}/{\rm
  dt}=(-0.70\pm 0.28)\cdot10^{-8}$\,s/s, and the corresponding
$\tau_{\rm sync}=153^{+102}_{-44}$\,yr have values in between the
results for the analysis of the optical data and the combined analysis
of optical and X-ray data sets.

Therefore, it is most likely that the discrepancy between the values
for ${\rm dP}_{\rm spin}/{\rm dt}$ and $\tau_{\rm sync}$ derived from
the optical data set and the combined optical and X-ray data is caused
by a systematic shift between the trough timings in both wavebands.
Thus, we judge the values for ${\rm dP}_{\rm spin}/{\rm dt}$ for
\object{V1432\,Aql} from the analysis of the coherent optical data set
as the more reliable and appropriate estimates of this parameter,
yielding a synchronization time scale $\tau_{\rm sync}$ of
approximately 200\,yr. Still, the underlying physical cause for the
troughs (absorption in the accretion funnel directly above the
accretion region) is the same. Because the X-ray data precede the
optical data (Fig.~\ref{aspinres}), a slightly variable ${\rm dP}_{\rm
  spin}/{\rm dt}$ (and therefore inclusion of terms of higher order
into the ephemeris) also might contribute to the discrepancy.

The time scale $\tau_{\rm sync}$ for the synchronization in
near-synchronous systems can be calculated by estimating the torques
acting on the white dwarf by the infalling matter and the magnetic
interaction of the binary components (Hameury et al., 1989,
Campbell \& Schwope, 1999). Using relations~(8) and~(10)
from Warner \& Wickrama\-singhe~(1991) and adapting them for
near-synchronous systems, we obtain:

\begin{eqnarray}
\frac{\tau_{\rm sync}^{\rm acc}}{1.1\times 10^4\,{\rm y}}&=&\left(
\frac{{\rm P}_{\rm beat}}{\rm d}\right)^{-1}
\left(\frac{{\rm P}_{\rm orb}}{\rm 4\,h}\right)^{-1/3}
\left(\frac{\dot {\rm M}}{\rm 10^{17}\,g/s}\right)^{-1}\\
\frac{\tau_{\rm sync}^{\rm mag}}{4.1\times 10^3\,{\rm y}}&=&\left(
\frac{{\rm P}_{\rm beat}}{\rm d}\right)^{-1}
\left(\frac{{\rm P}_{\rm orb}}{\rm 4\,h}\right)^2
\left(\frac{\mu_{\rm wd}\,\mu_{\rm sec}}{\rm 10^{68}\,G^2\,cm^6}\right)^{-1}
\end{eqnarray}
with the magnetic moments $\mu_{\rm wd}$ of the white dwarf and
$\mu_{\rm sec}$ of the secondary (intrinsic plus induced components),
the mass accretion rate $\dot {\rm M}$ and with the mass of the white
dwarf set to ${\rm M}_{\rm wd}=0.6\,{\rm M}_\odot$.

In the unique system \object{V1432\,Aql} with ${\rm P}_{\rm spin}>{\rm
  P}_{\rm orb}$ both torques on the white dwarf by the infalling
matter and the magnetic interaction with the secondary are acting in
concert towards a synchronization of the white dwarf spin period with
the orbital period, resulting in a relatively short synchronization
time scale. In the other near-synchronous systems \object{BY\,Cam},
\object{V1500\,Cyg} and \object{RX\,J2115--58}, only the magnetic
interaction will work towards a (re-) synchronization of the white
dwarf spin rotation with the binary period.

Using the ${\rm P}_{\rm orb}$--$\dot {\rm M}$ relation for cataclysmic
variables ($\dot {\rm M}\propto {\rm P}_{\rm orb}^{3.2}$,
Patterson, 1984 - as already used in Section 6) 
we obtain an estimate of the mass accretion
rate for \object{V1432\,Aql} of $\dot {\rm M}=2\cdot 10^{16}$\,g/s.
The resulting synchronization time scale from the torque of the
accreted matter is $\tau^{\rm acc}_{\rm sync}\approx10^3$\,yr.  With
typical magnetic moments for both binary components ($\mu_{\rm
  wd}=1\cdot 10^{34}$\,G\,cm$^3$ for a white dwarf primary with ${\rm
  M}_{\rm wd}=0.6\,{\rm M}_\odot$ and ${\rm B}_{\rm pol}=30$\,MG;
$\mu_{\rm sec}=2\cdot 10^{34}$\,G\,cm$^3$ for the secondary, see
Warner \& Wickramasinghe, 1991) the resulting synchronization
time scale from the magnetic torque is $\tau^{\rm mag}_{\rm
  sync}\approx30$\,yr.  The magnetic torque thus dominates the
synchronization process in the near-synchronous system
\object{V1432\,Aql}. Generally it is assumed that white dwarfs in
polars are more massive. For a white dwarf with ${\rm M}_{\rm wd}=
1.0\,{\rm M}_\odot$ ($\mu_{\rm wd}=3\cdot 10^{33}$\,G\,cm$^3$ with
${\rm B}_{\rm pol}=30$\,MG) we obtain $\tau^{\rm mag}_{\rm
  sync}\approx200$\,yr.

The synchronization time scale $\tau_{\rm sync}$ observed in
\object{V1432\,Aql} thus can be theoretically explained by the action
of the dominating magnetic torque on the white dwarf due to the
interaction of the magnetic fields of both binary components. If we
assume that \object{V1432\,Aql} would be an intermediate polar with
${\rm P}_{\rm spin}\approx4040$\,s, as proposed by
Mukai~(1998), the observed ${\rm dP}_{\rm spin}/{\rm dt}$
would be approximately two orders of magnitude higher than typical
values for Intermediate Polars (Patterson, 1994). Therefore our
results for the
synchronization time scale favor the interpretation that
\object{V1432\,Aql} is indeed a near-synchronous system, albeit the
only one known with ${\rm P}_{\rm spin}>{\rm P}_{\rm orb}$, which
places an interesting theoretical problem and is not yet understood.

\section{Conclusions}

Systematic studies of near-synchronous polars extending over at least
one beat period provide a unique opportunity to gain insight into the
not well understood interaction between the infalling accretion stream
and the white dwarf magnetosphere. This allows to find tighter 
constraints on model parameters as in the case
of locked polars.

We have presented a dipole accretion model for the description of the
accretion process in near-synchronous systems in general and
demonstrated its applicability to \object{V1432\,Aql}. We have
demonstrated how the model allows to constrain the accretion geometry
in this system.

We have observed a secular change of the white dwarf spin period in
\object{V1432\,Aql}, resulting in a synchronization time scale of
$\sim200$\,yr, comparable to the time scale for the white dwarf in
\object{V1500\,Cyg}. After \object{V1500\,Cyg} and \object{BY\,Cam},
\object{V1432\,Aql} is the third near-synchronous system (out of four)
with a known secular change of the white dwarf spin period.  The
synchronization time scale is in excellent agreement with theoretical
predictions from the torques in near-synchronous systems.

We put forward arguments against the alternative model of the system
\object{V1432\,Aql} as an intermediate polar with ${\rm P}_{\rm
  spin}\approx4040$\,s (Mukai, 1998), and for the
interpretation of this peculiar system as the only near-synchronous
polar with ${\rm P}_{\rm spin}>{\rm P}_{\rm orb}$.

\acknowledgements{This work is largely based on the doctoral thesis
of R.D.\ Geckeler, submitted to the Physics Faculty of the University 
of T\"ubingen in 1998, and an unpublished manuscript originally 
submitted to A\&A by Geckeler et al. in 2000. 
We thank the referee for a very detailed
and constructive report which allowed to improve the paper significantly.
This work had been supported by DARA grant 50009605 and FWF grant 
P\,11675--AST.  R.\ Rothschild acknowledges support by NASA contract
NAS5--30720 and NSF grant INT--9815741.}

\end{document}